\documentclass[sigconf]{acmart}

\usepackage{booktabs}
\usepackage{tabularx}
\usepackage{multirow}
\usepackage{multicol}
\usepackage{balance}
\usepackage[utf8]{inputenc}
\usepackage{kotex}
\usepackage{enumitem}
\usepackage[ruled,lined, linesnumbered]{algorithm2e}
\usepackage{algpseudocode}
\usepackage{svg}
\usepackage{graphicx}

\usepackage{lipsum}

\newlength\mylen
\newcommand\myinput[1]{%
  \settowidth\mylen{\KwIn{}}%
  \setlength\hangindent{\mylen}%
  \hspace*{0.5\mylen}#1\\}
\usepackage{tabto}

\newcommand{\itab}[1]{\hspace{0em}\rlap{#1}}
\newcommand{\ttab}[1]{\hspace{.07\textwidth}\rlap{#1}}
\let\oldnl\nl
\newcommand{\nonl}{\renewcommand{\nl}{\let\nl\oldnl}}
\newcolumntype{P}[1]{>{\centering\arraybackslash}p{#1}}

\setcopyright{acmcopyright}

\acmDOI{10.1145/3672608.3707958}

\acmISBN{978-1-4503-8713-2/22/04}

\acmConference[SAC'25]{ACM SAC Conference}{March 31 - April 4, 2025}{Sicily, Italy}  
\acmYear{2025}
\copyrightyear{2025}

\begin{document}
\title[]{Topic-Aware Knowledge Graph with Large Language Models for Interoperability in Recommender Systems}

\author{Minhye Jeon}

\orcid{0009-0000-7957-4442}
\affiliation{%
  \institution{Inha University}
  \city{Incheon} 
  \state{South Korea} 
}
\email{push@inha.edu}

\author{Seokho Ahn}
\orcid{0000-0002-5715-4507}
\affiliation{%
  \institution{Inha University}
  \city{Incheon} 
  \state{South Korea} 
}
\email{sokho0514@inha.edu}

\author{Young-Duk Seo}
\authornote{ Corresponding author}
\orcid{0000-0001-8542-2058}
\affiliation{%
  \institution{Inha University}
  \city{Incheon} 
  \state{South Korea} 
}
\email{mysid88@inha.ac.kr}

\renewcommand{\shortauthors}{Jeon et al.}

\begin{abstract}
The use of knowledge graphs in recommender systems has become one of the common approaches to addressing data sparsity and cold start problems.
Recent advances in large language models (LLMs) offer new possibilities for processing side and context information within knowledge graphs. 
However, consistent integration across various systems remains challenging due to the need for domain expert intervention and differences in system characteristics.
To address these issues, we propose a consistent approach that extracts both general and specific topics from both side and context information using LLMs.
First, general topics are iteratively extracted and updated from side information. 
Then, specific topics are extracted using context information.
Finally, to address synonymous topics generated during the specific topic extraction process, a refining algorithm processes and resolves these issues effectively.
This approach allows general topics to capture broad knowledge across diverse item characteristics, while specific topics emphasize detailed attributes, providing a more comprehensive understanding of the semantic features of items and the preferences of users.
Experimental results demonstrate significant improvements in recommendation performance across diverse knowledge graphs.

\end{abstract}



\maketitle

\section{Introduction}
\label{sec:01_introduction}
Recommender systems mainly rely on direct interactions between users and items, which often leads to challenges such as data sparsity and cold start problems. 
To address these challenges, the side information of items (\textit{e.g.}, price, category, and brand) offers an opportunity to leverage diverse relations and information.
In particular, the diverse characteristics of recommender systems and domain-specific differences make it challenging to construct and apply a knowledge graph consistently across systems.
In response, previous studies \cite{lee2023relation, kwon2024reckg} have highlighted the importance of interoperability in knowledge graph-based recommender systems by standardizing the structure of side information or consistently modeling relations.

\begin{figure}
    \centering
    \vspace{2em}
    \includegraphics[width=1\columnwidth]{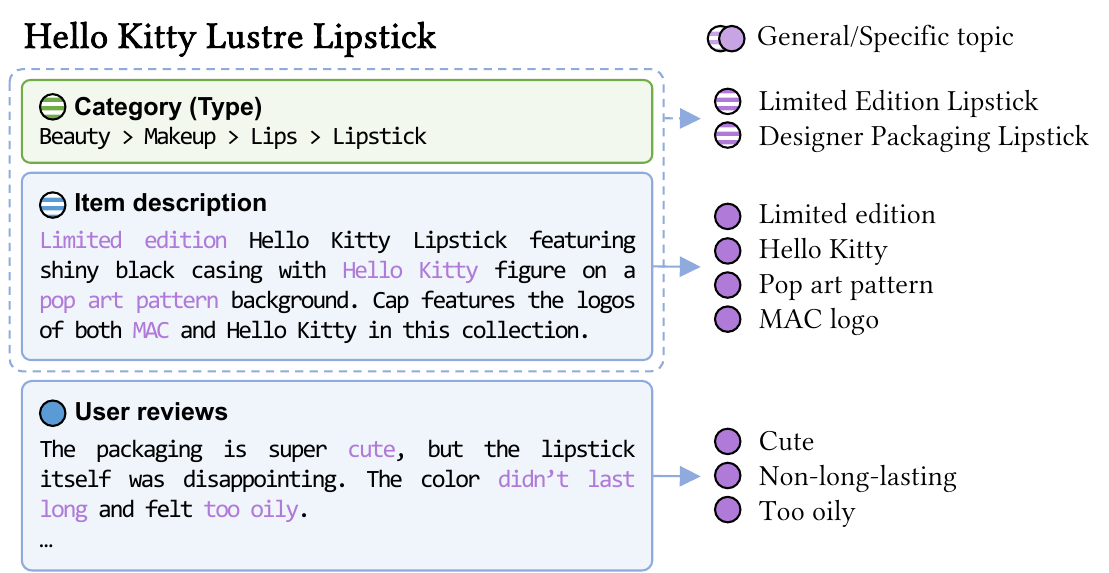}
    \caption{Visualization of general and specific topic extraction \textnormal{from structured side information and unstructured context information. }}
    \label{fig:enter-label}
\end{figure}

Recently, large language models (LLMs) have demonstrated their ability to deeply understand and infer meaning from context information across various domains, driving research into methods for extracting keywords from context information. 
Semantic features within context information (\textit{e.g.,} item descriptions and user reviews) offer valuable insights but are often utilized without proper processing or analysis in knowledge graph-based recommender systems.
However, applying these methods to recommender systems that prioritize interoperability remains challenging, particularly due to the reliance on for domain experts to manually integrate context information into knowledge graphs. 
Additionally, the methods used to expand these knowledge graphs are often inconsistent, making it difficult to systematically organize the diverse information embedded in both side and context information.
Finally, identifying and processing synonymous words generated by LLMs requires additional steps, which need to be handled consistently.

To address this problem, we propose a systematic and consistent method for extracting topics inherent in the side and context information of items, leveraging LLMs. 
Building on the need to comprehend knowledge across multiple contextual levels \cite{crump2020generalspecific}, our approach extracts both general and specific topics to capture the semantic features of the items from various perspectives. 
Figure \ref{fig:enter-label} illustrates an example of the results from our topic extraction process.
The proposed method involves three key steps: \textsf{\textit{(i) General topic extraction}}. Using both side and context information, we extract a general topic. 
This step addresses overlapping data by incorporating the previously extracted general topic into subsequent iterations, ensuring consistency across the process;
\textsf{\textit{(ii) Specific topic extraction}}. Context information is utilized to extract specific topics, capturing the unique characteristics of each item; \textsf{\textit{(iii) Topic refinement}}. A refining algorithm is applied to address synonym overlap within the specific topics, ensuring clarity and consistency despite their inherent complexity;

To achieve this, the general topic offers deeper insights into broader characteristics (\textit{e.g.}, sub-category) compared to the original side information (\textit{e.g.}, category).
In contrast, the specific topic captures each item's information and users' preferences, reflecting both subjective and objective aspects.
Leveraging both the general and specific topics enables a more comprehensive understanding of the characteristics of an item, as well as the preferences and behaviors of its users.
Additionally, our approach is based on a standardized metagraph, ensuring a consistent and interoperable approach across various domains in recommender systems.
We demonstrate that the proposed approach significantly improves recommendation performance by comparing results across diverse knowledge graphs in multiple systems.

The main contributions are summarized as follows:

\begin{itemize}[leftmargin=1.1em]
\item We propose a method that extracts general and specific topics from both side and context information within the knowledge graph using LLMs.
\item Our approach ensures interoperability by utilizing a standardized metagraph, enabling consistent topic extraction and refining processes regardless of the differences across various systems.
\item Comparative experiments demonstrate that the proposed approach improves recommendation performance in terms of four different evaluation metrics across various domains.
\end{itemize}

\section{Related Works}
\label{sec:02_relatedworks}

This section first explains previous work on knowledge graph-based recommender systems (Section \ref{sec:02_KG_RS}), and then introduces previous studies that enhance knowledge graphs in recommender systems using LLMs (Section \ref{sec:02_LLM_RS}).

\subsection{Knowledge graph-based Recommender Systems} \label{sec:02_KG_RS}

Approaches to leveraging knowledge graphs in recommender systems are generally categorized into two types: \textsf{\textit{(i) Embedding-based approach}} and \textsf{\textit{(ii) Path-based approach}}.
These approaches differ in how they leverage the relations within the knowledge graph.
The embedding-based approach represents entities and relations by embedding them into a low-dimensional vector space, enabling recommendations based on vector similarity. 
In contrast, the path-based approach identifies connection paths between users and items, utilizing these paths for recommendations. 
This approach allows for discovering indirect connections through path exploration, enabling recommendations even when direct links are latent.
We provide a detailed explanation of each approach as follows.

\subsubsection*{Embedding-based approach. }
\label{sec:02_embedding-based}
TransE \cite{bordes2013transe} and TransR \cite{lin2015transr} are among the most widely used techniques for knowledge graph embedding.
CKE \cite{zhang2016cke} is a hybrid model that integrates collaborative filtering with TransR to capture the latent relations between users and items. 
KGAT \cite{wang2019kgat} and CKAN \cite{wang2020ckan} extend TransR-based embeddings by incorporating an attention mechanism to assign varying importance to relations, thereby enhancing model performance. 
RippleNet \cite{wang2018ripplenet} and KGIN \cite{wang2021kgin} both leverage TransE for knowledge graph embeddings. 
RippleNet focuses on propagating user preferences, while KGIN learns the relational characteristics to enhance the latent relations between users and items.
However, embedding-based approaches are constrained to utilize the detailed characteristics of items, as they primarily focus on structural information \cite{zou2020kgsurvey}.

\subsubsection*{Path-based approach. }
Designing meta-paths or modeling inter-entity connection patterns presents a significant challenge \cite{guo2020kgrssurvey}. 
KPRN \cite{wang2018kprn} extracts various paths connecting users and items from knowledge graphs and learns them using recurrent neural networks (RNNs). 
RuleRec \cite{ma2019rulerec} identifies path rules from knowledge graphs to capture latent relations, learning rules that users tend to follow. 
PGPR \cite{xian2019pgpr} and ReMR \cite{wang2022remr} employ reinforcement learning to discover user-item paths. 
However, when knowledge graphs lack sufficient paths between users and items, the amount of information available for the model to learn is limited, leading to potential performance degradation. 
Despite this, previous work has not consistently leveraged context information to address these limitations. 
Since relevant information varies across domains, some domains require expertise from domain experts \cite{tufis2014domainexpertknowledge}, which leads to challenges in terms of scalability and generalizability.

\subsubsection*{Our contributions. }
To address this problem, we propose a method to effectively utilize context information in a knowledge graph built from the side information of items. 
We also present a consistent approach to enhancing the knowledge graph.
Furthermore, we demonstrate that LLMs can be applied without expert intervention, improving interoperability.

\subsection{LLM-based Knowledge graph Construction in Recommender Systems} \label{sec:02_LLM_RS}
Recently, research on enhancing knowledge graphs using LLMs has been actively progressing. 
Previous studies aim to address the data sparsity and cold start problems in recommender systems by expanding entities and relations using LLMs \cite{zhao2024breaking, jiang2024diffkg}.

\citeauthor{yang2024sequential} \cite{yang2024sequential} proposes a framework that utilizes side information to discover latent relations, suggesting a method to automatically identify new types of relations using LLMs to enhance recommendation performance.
\citeauthor{shi2024llmsubgraph} \cite{shi2024llmsubgraph} extracts qualitative information, such as style, price, and color, from user review data to build user subgraphs, improving recommendation performance through subgraph inference. 
\citeauthor{yang2024commonsensellm} \cite{yang2024commonsensellm} introduces a method that generates common sense-based subgraphs from item side information using LLMs and applies them to recommendations. 
However, previous studies often do not fully utilize context information, relying only on limited side information such as type and brand, or extracting only general characteristics of items by leveraging external knowledge from LLMs \cite{yang2024commonsensellm}.

\subsubsection*{Our contributions. }
In this paper, we extract more enriched general topics compared to previous studies by utilizing not only side information but also context information, which captures item-specific semantic features.
Furthermore, we enhance the knowledge graph by extracting specific topics that capture the characteristics of each item and the preferences of users based on context information.

\section{Preliminaries}
\label{sec:03_preliminaries}

This section first introduces the theoretical backgrounds (Section \ref{sec:03_preliminaries_theoretical}) and formulates the task (Section \ref{sec:03_preliminaries_formulation}) of our proposed approach.

\subsection{Theoretical Backgrounds} \label{sec:03_preliminaries_theoretical}
This section provides theoretical explanations of three key concepts essential for understanding the proposed approach: \textsf{\textit{Knowledge graph}}, \textsf{\textit{Metagraph}}, and \textsf{\textit{Side/Context information}}.

\subsubsection*{Knowledge graph}

A knowledge graph is a directed graph consisting of nodes and edges, where each edge represents a semantic relationship between two nodes \cite{peng2023kgchallenges, lee2023relation}. We formally define a knowledge graph \(\mathcal{G}\) as: 
\begin{equation}
\mathcal{G}=\left\{\left(h, r, t\right) \mid h, t \in \mathcal{E}, r \in \mathcal{R}\right\}
\end{equation}
where \(\mathcal{E}=\left\{e_1, e_2, \cdots, e_k\right\}\) is a set of \(k\) entities (\textit{i.e.}, nodes) and \(\mathcal{R}=\left\{r_1, r_2, \cdots, r_g\right\}\) is a set of \(g\) relation types (\textit{i.e.}, edges). In this context, \(h, r, \) and \(t\) typically denote the head, relation, and tail, respectively. Each triplet \(\left(h, r, t\right)\in\mathcal{G}\) indicates a semantic relationship \(r\) from the entity \(h\) to the entity \(t\).

In recommender systems, the entity types can include users \(\mathcal{U}\) and items \(\mathcal{I}\) in addition to knowledge base entities \(\mathcal{E}\). Incorporating explicit and implicit feedback between users and items, such as \(\mathcal{R^+}=\left\{\textsf{interact}, \textsf{also\_bought}, \cdots \right\}\), can provide more insights into user preferences \cite{sun2018rkge, wang2018kprn}.
Additionally, newly defined relations between other entities can be considered \cite{lee2023relation}. To capture all these interactions, we can expand the knowledge graph \(\mathcal{G}\) to an extended knowledge graph \({\mathcal{G}_\textrm{RS}}=\{(h, r, t)\mid h, t \in \mathcal{E}\cup\mathcal{U}, r \in\mathcal{R}\cup\mathcal{R}^+\}\) for recommender systems.

\begin{figure*}[tbh!]
\includegraphics[width=1\textwidth]{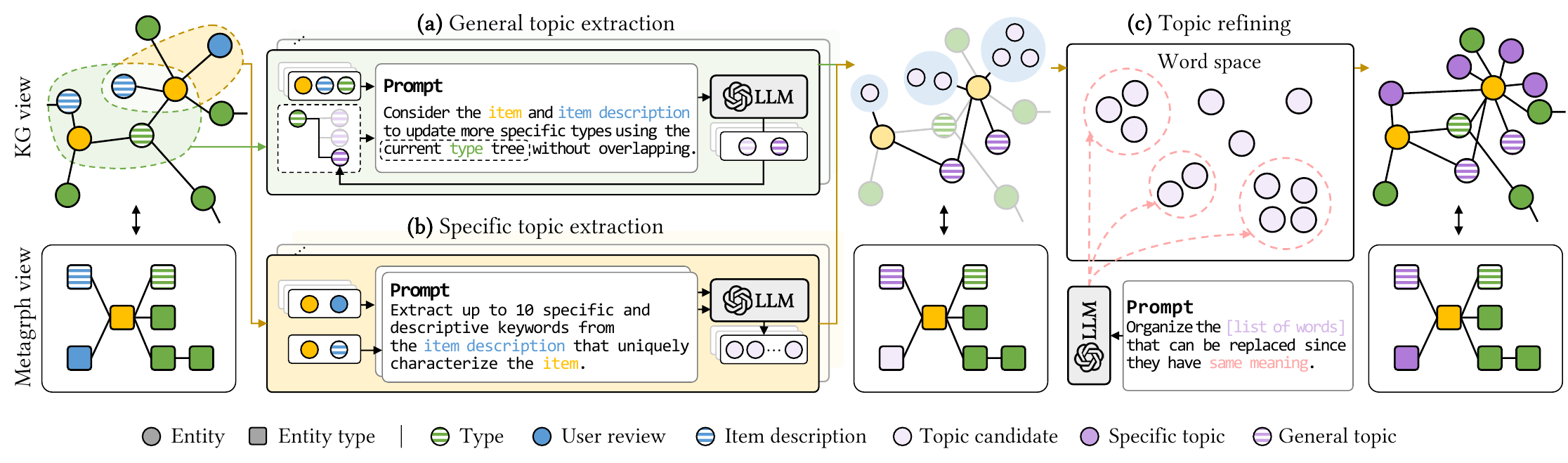}
\centering
\caption{Overview of our proposed approach. \normalfont{Each color represents an entity in the graph: \textcolor[rgb]{1, 0.753, 0}{\textsf{\textit{\textbf{yellow}}}} for items, \textcolor[rgb]{0.439, 0.678, 0.278}{\textsf{\textit{\textbf{green}}}} for side information, \textcolor[rgb]{0, 0.427, 0.85}{\textsf{\textit{\textbf{blue}}}} for context information, and \textcolor[rgb]{0.604, 0.341, 0.804}{\textsf{\textit{\textbf{purple}}}} for topics extracted from the context information. This figure shows a summary of the task prompt; the full prompt is provided in Figure \ref{fig:prompt}.}}
\label{fig:framework}
\end{figure*}

\subsubsection*{Metagraph} 
We first define a metagraph \(\mathcal{M}\) of a knowledge graph \(\mathcal{G}\), also known as a \textit{schema} \cite{nicholson2020biomedicalmetagraph, ye2023schema}, which represents the relationships between super-entities \cite{chung2021metagraphlearning}:
\begin{equation}
\mathcal{M}=\left\{\left(\eta, r, \tau\right) \mid \eta, \tau \in \mathcal{A}, r \in \mathcal{R}\right\}
\end{equation}
where \(\mathcal{A}\) denotes a set of entity types. 
The metagraph 
provides additional abstract information about entities \cite{lu2024schemaaware} but also 
enforces the triplet types of the knowledge graph \cite{kwon2024reckg}, thereby enhancing the interoperability of different systems. 
Formally, we first define the subset of entities \(\mathcal{E}_a=\left\{e\in\mathcal{E}\mid\phi(e)=a\right\}\) that corresponds to an entity type \(a\in\mathcal{A}\), using an entity type mapping function \(\phi:\mathcal{E}\rightarrow\mathcal{A}\).
Then for a triplet \((h, r, t)\in\mathcal{G}\), there \textit{must} exist a triplet type \((\eta, r, \tau)\in\mathcal{M}\) where \(h\in\mathcal{E}_\eta\) and \(t\in\mathcal{E}_\tau\).

For recommender systems, constructing knowledge graphs in a consistent manner is challenging due to their variations in domains and system characteristics.
As part of this effort, RecKG \cite{kwon2024reckg} standardizes the metagraph \(\mathcal{M}^*\) for recommender systems by categorizing entity types  \({\mathcal{A}^*}\), relation types \({\mathcal{R}^*}\), and their triplet types.
By constructing knowledge graphs based on this standardized metagraph, multiple systems can be integrated more effectively, thus achieving interoperability.
In other words, expanding a standardized metagraph provides a basis for developing improved methods that achieve interoperability.

\subsubsection*{Side/Context information}
We begin by grouping the standardized entity types \(\mathcal{A}^*\) within the standardized metagraph \(\mathcal{M}^*\) for recommender systems, aiming to develop an improved method while maintaining interoperability.
Based on the metagraph proposed in RecKG \cite{kwon2024reckg}, the entity types include a user \(\mathcal{U}\in\mathcal{A^*}\) and an item \(\mathcal{I}\in\mathcal{A}^*\), while the remaining entity types provide supplementary information for these users and items.

We first formally define the type of \textit{side information} \cite{sattar2023sideinfo, guo2020kgrssurvey} as a subset of all entity types 
except for the user and item, \(\mathcal{A}_\textrm{side}=\mathcal{A}^*-\left\{\mathcal{U}, \mathcal{I}\right\}\).
Examples of side information include attributes such as \textsf{Performer} (a standardized form for roles such as actor and singer), \textsf{Type} (a standardized form for classifications such as category and genre), \textsf{Release date}, \textsf{Price}, and \textsf{Description}.
Assuming the head entity type is either a user \(\mathcal{U}\) or an item \(\mathcal{I}\),
we also define the set of relation types where the tail entity type is side information, \textit{i.e.}, \(\mathcal{R}_\textrm{side}=\{r \in\mathcal{R} \mid \left(\eta, r, \tau\right)\in\mathcal{M}^*,  \tau \in\mathcal{A}_\textrm{side}\}\).

Among the side information, we focus on textual data as \textit{context information} \cite{sattar2023sideinfo, sattar2021contextinfo} due to its ability to capture richer contextual meanings beyond categorical or numerical attributes, such as \textsf{Description} and \textsf{Review}.
Formally, let \(\mathcal{A}_\textrm{cont}\subseteq\mathcal{A}_\textrm{side}\) represent the set of entity types that provide context information, and
\(\mathcal{R}_\textrm{cont}=\{r \in\mathcal{R} \mid \left(\eta, r, \tau\right)\in\mathcal{M}^*,  \tau\in\mathcal{A}_\textrm{cont}\}\) denote the corresponding set of contextual relation types.
These contextual attributes are particularly valuable in recommender systems as they enable the extraction of semantic features through LLMs \cite{zheng2017texttosemantic, wang2024textinfopretrained}, allowing for more nuanced and personalized recommendations while maintaining interoperability.

\subsection{Task Formulation \label{sec:03_preliminaries_formulation}}

This section outlines the objectives of our approach. 
Specifically, we aim to construct a topic-aware knowledge graph \(\overline{\mathcal{G}}\) (output) from an existing knowledge graph \(\mathcal{G}_\textrm{RS}\) (input) by consistently extracting and replacing topics within context information. 
To ensure interoperability across different recommender systems, we begin by constructing a topic-aware metagraph \(\overline{\mathcal{M}}\) based on a standardized metagraph \(\mathcal{M}^*\). 
Formally, we first define a subset \({\mathcal{M}_\textrm{base}} \subseteq \mathcal{M}^{*}\), containing all triplet types except for context information, as follows:
\begin{equation}
\mathcal{M}_\textrm{base} =\left\{\left(\eta, r, \tau\right)\mid \eta, \tau \in \mathcal{A}^{*}-\mathcal{A}_\textrm{cont}, r \in \mathcal{R}^{*}-\mathcal{R}_\textrm{cont}\right\}.
\end{equation}
Next, we extract the topic entity type \(\mathcal{A}_\textrm{topic}\) from both side and context information, along with its corresponding relation \(\mathcal{R}_\textrm{topic}\). 
Constructing topic-only metagraph \(\mathcal{M}_\textrm{topic}\) based on these topic entity types \(\mathcal{A}_\textrm{topic}\) and its relations \(\mathcal{R}_\textrm{topic}\) as follows: 
\begin{equation}
\mathcal{M}_\textrm{topic} = \left\{\left(\eta, r, \tau\right) \mid \eta \in \left\{\mathcal{U}, \mathcal{I}\right\}, r \in \mathcal{R}_\textrm{topic}, \tau \in \mathcal{A}_\textrm{topic}\right\}.
\end{equation}
Then the final topic-aware metagraph \(\overline{\mathcal{M}}\) is constructed by combining the base metagraph \(\mathcal{M}_\textrm{base}\) and topic-only metagraph \(\mathcal{M}_\textrm{topic}\).
This union integrates the fixed side information with the newly extracted topic information:
\begin{equation}
\overline{\mathcal{M}} = \mathcal{M}_\textrm{base} \cup \mathcal{M}_\textrm{topic}.
\end{equation}
This combination ensures that all contextual properties are replaced by the corresponding topics, eliminating context information from the metagraph.
As the topic-aware metagraph \(\overline{\mathcal{M}}\) is derived from the standardized metagraph \(\mathcal{M}^*\), interoperability is ensured without difficulty.
Finally, enforcing triplet types \(\left(\eta, r, \tau\right)\in\overline{\mathcal{M}}\), we consistently construct the topic-aware knowledge graph \(\overline{\mathcal{G}}\):
\begin{equation}
\overline{\mathcal{G}}=\left\{\left(h, r, t\right) \mid \left(\eta, r, \tau\right) \in \overline{\mathcal{M}}, h \in \mathcal{E}_\eta, t \in \mathcal{E}_\tau\right\}
\end{equation}
which achieves the aim of the task.

Note that there are several factors to consider before proceeding with this task. 
First, it is essential to determine which topic type to extract. 
Second, the process of extracting topics from the actual context information must be performed in a systematic and consistent manner.
The method proposed in the next section outlines various approaches for addressing these factors.

\begin{figure*}[t!]

\includegraphics[width=1\textwidth]{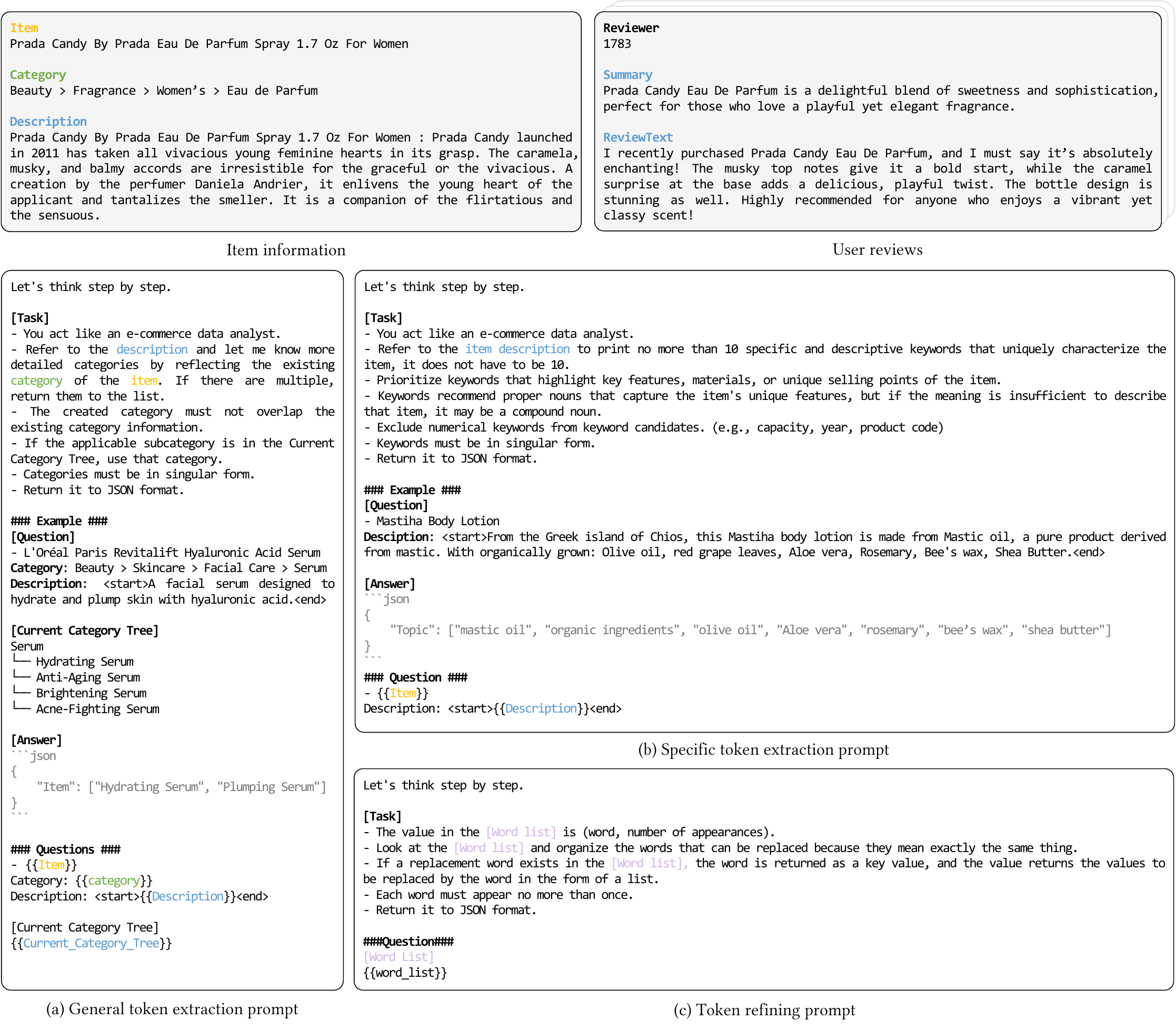}

\centering
\vspace{-0.7em}
\caption{Examples of item information and user reviews used in the proposed approach and corresponding prompts.}

\label{fig:prompt}
\end{figure*}

\section{Proposed Approach}
\label{sec:04_model}

This section first provides an overview of our approach (Section \ref{sec:04_model_overview}),
then details the extraction of general and specific topics from both side and context information in the knowledge graph (Sections \ref{sec:04_model_general} and \ref{sec:04_model_specific}). Finally, we focus on synonymous topics in the extraction process and introduce a refining algorithm to resolve this issue while maintaining interoperability (Section \ref{sec:04_model_refine}).

\subsection{Approach Overview \label{sec:04_model_overview}}

This section provides an overview of our proposed approach, shown in Figure \ref{fig:framework}.
As discussed in Section \ref{sec:03_preliminaries_formulation}, our approach aims to construct a topic-aware knowledge graph for recommender systems by extracting topic entities from both side and context information, ensuring interoperability.
Considering the importance of understanding knowledge across multiple contextual levels \cite{crump2020generalspecific}, we use distinct strategies for topic extraction using LLMs.
Specifically, we \textit{update} the general topics, which already exist in the side information (\textit{e.g}., \textsf{Type}); while we \textit{replace} the specific topics from the context information (\textit{e.g}., \textsf{Description} and \textsf{Review}).

To achieve this, we first extract the general topic using both side and context information, then update more detailed type (i.e. \textsf{SubType}) for lowest level \textsf{Type} entity (in Figure \ref{fig:framework}(a) and Section \ref{sec:04_model_general}).
Simultaneously, specific topics are replaced using context information, including \textsf{Description} and \textsf{Review}, for each \textsf{Item} entity (in Figure \ref{fig:framework}(b) and Section \ref{sec:04_model_specific}).
Due to the different extract strategies (\textit{i.e.}, \textit{update} and \textit{replace} strategies), specific topics may lead to synonymous entities across different items. To address this, we introduce a refining algorithm to handle these synonyms (in Figure \ref{fig:framework}(c) and Section \ref{sec:04_model_refine}).
By starting with a standardized metagraph, the entire process can be consistently applied across various recommender systems.
Each process is detailed in the following sections.

\subsection{General topic extraction \label{sec:04_model_general}}
This section describes the method for extracting general topics from the knowledge base, shown in Figure \ref{fig:framework}(a).
Focusing solely on \textsf{Type}) (\textit{e.g.}, category and genre) is too broad to capture the detailed categorization of an item. To address this, we enhance the \textsf{Type} entity by extracting a general topic.
Specifically, we use both side information (\textit{i.e.}, \textsf{Type}) and context information (\textit{i.e.}, 
 \textsf{Description}) to extract the \textsf{Subtype} entity.

Formally, a triplet type is first added to the topic-aware metagraph \(\overline{\mathcal{M}}\) to ensure interoperability, satisfying:
\begin{equation}
\{(\mathcal{I}, \textsf{related\_to}, \textsf{Subtype})\} \subseteq \overline{\mathcal{M}}.
\end{equation}
It follows that \(\textsf{Subtype} \in \overline{\mathcal{A}}\) and \(\textsf{related\_to} \in \overline{\mathcal{R}}\) hold.
By fixing these triplet types, we construct entities \(\mathcal{E}_\textsf{Subtype}\subseteq\overline{\mathcal{E}}\) by using LLMs based on both \textsf{Type} and \textsf{Description} entity types.
These triplets are added to the topic-aware knowledge graph \(\overline{\mathcal{G}}\):
\begin{equation}
\left\{\left(h, r, t\right) \mid h \in \mathcal{I}, r=\textsf{related\_to}, t \in \mathcal{E}_\textsf{Subtype}\right\}\subseteq \overline{\mathcal{G}}.
\end{equation}

During the construction of \(\mathcal{E}_\textsf{Subtype}\), it is iteratively updated for each leaf node in the type tree, which organizes the existing \textsf{Type} elements in the knowledge base into a tree structure, starting from \(\mathcal{E}_\textsf{Type}\).
This process is iterated for items within the same lowest level \textsf{Type} in the type tree.
We provide an illustrative example in Figure \ref{fig:prompt}(a) to explain this iterative process:

The general topic reflects the information from the \textsf{\textit{``Current Category Tree''}}, which contains subtypes generated from items at the same lowest level. 
Leveraging existing data, the subtype of the current item is added to the \textsf{\textit{``Current Category Tree''}}, if it is not duplicate.
For instance, when \textsf{\textit{``Hydrating Serum''}} and \textsf{\textit{``Plumping Serum''}} are generated, the non-duplicate subtypes are added to the \textsf{\textit{``Current Category Tree''}}.
By repeating this process for each item, the extraction of general topics effectively captures the detailed information compared to the existing \textsf{Type}.

\subsection{Specific topic extraction \label{sec:04_model_specific}}
This section focuses on extracting specific topics, as shown in Figure \ref{fig:framework}(b).
Specific topics represent unique properties inherent in context information that cannot be derived from side information. 
For example, \textsf{Description} provides objective information about an item, whereas \textsf{Review} conveys subjective opinions from users who have interacted with the item.

To achieve this, we extract and replace inherent topics within such context information, iterating through all connected context information for each item.
In this context, each item is assumed to have \(N\) reviews and a single description, as shown in the left part of Figure \ref{fig:framework}(b). 
Specific topics (\textit{i.e.}, \textsf{Word}) are extracted for each context information. 
From the \textsf{Review}, two triplets are added to the topic-aware metagraph
\(\overline{\mathcal{M}}\): 
\begin{equation}
    \{(\mathcal{U}, \textsf{mention}, \textsf{Word}), (\mathcal{I}, \textsf{described\_as}, \textsf{Word})\} \subseteq \overline{\mathcal{M}}.
\end{equation}
Similarly, a triplet type related to the \textsf{Description} is also added:
\begin{equation}
    \{(\mathcal{I}, \textsf{tagged}, \textsf{Word})\} \subseteq \overline{\mathcal{M}}.
\end{equation}
As shown in Figure \ref{fig:prompt}(b), the process extracts specific topics (i.e., \textsf{Word}) for each context information.
While \textsf{Word} extracted from \textsf{Description} contains objective information about the item, the information from \textsf{Review} reflects the subjective opinions of each review, thus distinguishing the relations between entities.

Note that this process is applied uniformly to all context information associated with an item.
Thus, unlike general topic extraction, the large number of candidate words leads to inherent synonyms within the entities \(\mathcal{E}_\textsf{Word}\).
To address this issue, the words are first aggregated as \(\textsf{Candidate\_word}\) and then grouped under \(\mathcal{E}_\textsf{Candidate\_word}\), requiring an additional synonym refinement process.
The following section provides a detailed explanation of this refinement process.

\subsection{Topic Refining \label{sec:04_model_refine}}
This section introduces an algorithm for refining specific topics, as illustrated in Figure \ref{fig:framework}(c).
As mentioned in Section \ref{sec:04_model_specific}, \(\mathcal{E}_\textsf{Candidate\_word}\) may have overlapping meanings.
Specific topics capture detailed characteristics of an item, often requiring a large number of words to fully describe the item's attributes.
However, using LLMs to handle synonyms for such a large number of specific topics within a single prompt, as in the method described in Section \ref{sec:04_model_general}, is challenging due to token limitations.
Therefore, an additional refining method is necessary to resolve this issue while maintaining interoperability.

Given the importance of both morphological and semantic similarity of words \cite{gladkova2016analogy}, we first partition the topics based on morphology to create manageable subsets for processing with LLMs, and subsequently group them semantically using LLMs.
To achieve the first step, we partition all extracted candidate topics \(\mathcal{E}_\textsf{Candidate\_word}\) as described in Algorithm \ref{alg:partition}. The detailed explanation of the algorithm is as follows:

This algorithm recursively partitions specific candidate words \(\mathcal{E}_\textsf{Candidate\_word}\), into smaller subsets based on their prefixes. 
Specifically, the function \(\texttt{TopicPartition}\) calls the recursive function \(\texttt{\_TopicPartitionRecursive}\), which returns partitions of  \(\mathcal{E}_\textsf{Candidate\_word}\) (Lines 1--4).
The function \texttt{\_TopicPartitionRecursive} takes two parameters: \(prefix\) and the subset of words \(W\).
Words whose first \(N\) characters match \(new\_prefix\) are added to the subset \(p\) (Lines 10--14).
If the size of \(p\) does not exceed the threshold \(T\), it is added to the partition \(\mathcal{P}\) (Lines 15--16). 
Otherwise, \texttt{\_TopicPartitionRecursive} is recursively called with the new prefix and the corresponding subset to further refine the partition (Lines 17--18).
The algorithm ensures that every subset in the final partition are smaller than \(T\).

Each subset \(p\) is individually fed into the LLMs to group \(\mathcal{E}_\textsf{Candidate\_word}\) with the same meaning, finally returning \(\mathcal{E}_\textsf{Word}\).
The prompt used in this process is shown in Figure \ref{fig:prompt}(c). 
In practice, the frequency of each topic's usage was included as input to the LLMs, replacing it with the more frequently mentioned topic.
Then these triplets are added to the topic-aware knowledge graph \(\overline{\mathcal{G}}\):
\begin{equation}
\begin{split}
& \left\{\left(h, r, t\right) \mid h \in \mathcal{U}, r=\textsf{mention}, t \in \mathcal{E}_\textsf{Word}\right\} \\
& \cup \left\{\left(h, r, t\right) \mid h \in \mathcal{I}, r=\textsf{described\_as}, t \in \mathcal{E}_\textsf{Word}\right\}  \\
& \cup \left\{\left(h, r, t\right) \mid h \in \mathcal{I}, r=\textsf{tagged}, t \in \mathcal{E}_\textsf{Word}\right\} \subseteq \overline{\mathcal{G}}.
\end{split}
\end{equation}
This process is not only constructed based on standardized metagraphs, but the algorithms are also applicable regardless of system size, ensuring interoperability.

\begin{algorithm}[t]
\caption{Specific Candidate Topic Partition}\label{alg:partition}
\DontPrintSemicolon
\SetKwInOut{Input}{Input}
\SetKwInOut{Output}{Output}

\Input{}
\nonl\myinput{\itab{\(\mathcal{E}_\textsf{Candidate\_word}\)} \ttab{\ttab{Specific candidate topic}}}
\nonl\myinput{\itab{\(T\)} \ttab{\ttab{Maximum subset size}}}

\Output{}

\nonl\myinput{\itab{\(\mathcal{P}\)} \ttab{\ttab{Partition of \(\mathcal{E}_\textsf{Candidate\_word}\)}}}
\nonl\;
\SetKwFunction{FMain}{TopicPartition} \label{alg:partition_start}
\SetKwProg{Fn}{Function}{:}{}
    \Fn{\FMain{}}{
        \(\mathcal{P} \longleftarrow \left\{\right\}\)

        \texttt{\_TopicPartitionRecursive}\(\left(\textnormal{""}, \mathcal{E}_\text{word}\right)\)

        \textbf{return} \(\mathcal{P}\) \label{alg:partition_end}
        }
\nonl\;
\SetKwFunction{FMain}{\_TopicPartitionRecursive}
\SetKwProg{Fn}{Function}{:}{}
    \Fn{\FMain{prefix, \(W\)}}{

        \textrm{Length} \(N \longleftarrow \textrm{len}(\textit{prefix})+1\)

        \ForEach{\textnormal{character} \(c \in\left[a\textrm{-}z, 0\textrm{-}9, \cdots\right]\)}
        {   
            
            \textrm{New prefix} \(\mathit{new\_prefix} \longleftarrow \textit{prefix} + c\) \label{alg:partition_nprefix}

            \textrm{Subset} \(p \longleftarrow \left\{\right\}\)

            \ForEach{\textnormal{word} \(w \in W\)} 
            {
                \If{\(w\left[0, N\right] = \mathit{new\_prefix}\)\label{line:dropout}}
                {\(p \longleftarrow p\cup \left\{w\right\}\)} 
            }

            \uIf{\(0< \textnormal{\textrm{size}}(p) \leq T\)}{
                \(\mathcal{P} \longleftarrow \mathcal{P}\cup \left\{p\right\}\)
              }
              \ElseIf{\(\textnormal{\textrm{size}}(p) > T\) \label{alg:partition_recursivestart}}{
                \texttt{\_TopicPartitionRecursive}\(\left(\mathit{new\_prefix}, p\right)\) \label{alg:partition_recursiveend}
              }
              }

        }
\end{algorithm}
\section{Experiments}
\label{sec:05_experiments}

In this section, we present our experimental setup and discuss the results of applying our proposed approach to the knowledge graph.

\begin{table*}[t]
\centering
\caption{Details of different knowledge graphs used in our experiment.}
\vspace{-0.6em}
\resizebox{0.88\textwidth}{!}{%
\setlength\tabcolsep{8pt}
\begin{tabular}{lcccccccc}
    \toprule
    \multicolumn{1}{l}{} & \multicolumn{4}{c}{\textbf{Amazon Beauty}}             & \multicolumn{4}{c}{\textbf{Amazon Clothing}}            \\
    \cmidrule(l{.3em}r{.3em}){2-5} \cmidrule(l{.3em}r{.3em}){6-9}
    \multicolumn{1}{l}{} & \textbf{\(\mathcal{G}_\textrm{base}\)}  & \textbf{\(\overline{\mathcal{G}}_\textrm{base}\) }& \textbf{\(\mathcal{G}_\textrm{large}\)}  & \textbf{\(\overline{\mathcal{G}}_\textrm{large}\)}  & \textbf{\(\mathcal{G}_\textrm{base}\)}  & \textbf{\(\overline{\mathcal{G}}_\textrm{base}\)} &\textbf{ \(\mathcal{G}_\textrm{large}\)}   & \textbf{\(\overline{\mathcal{G}}_\textrm{large}\) } \\
    \cmidrule(l{.3em}r{.3em}){1-1} \cmidrule(l{.3em}r{.3em}){2-3} \cmidrule(l{.3em}r{.3em}){4-5} \cmidrule(l{.3em}r{.3em}){6-7} \cmidrule(l{.3em}r{.3em}){8-9}
    \textbf{\#User}               & 22,363    & 22,363    & 22,363    & 22,363     & 39,387    & 39,387    & 39,387    & 39,387      \\ 
    \textbf{\#Item}               & 12,101    & 12,101    & 12,101    & 12,101     & 10,429    & 10,429    & 10,429     & 10,429     \\
    \textbf{\#Entity}             & 167,046   & 255,732   & 254,235   & 274,101    & 341,742   & 423,663   & 438,955    & 440,061    \\
    \textbf{\#General topic}      & -         & 9,933     & -         & 9,933      & -         & 553       & -          & 553        \\
    \textbf{\#Specific topic}     & -         & 78,753    & -         & 78,753     & -         & 81,368    & -          & 81,368     \\
    \cmidrule(l{.3em}r{.3em}){1-1} \cmidrule(l{.3em}r{.3em}){2-3} \cmidrule(l{.3em}r{.3em}){4-5} \cmidrule(l{.3em}r{.3em}){6-7} \cmidrule(l{.3em}r{.3em}){8-9}
    \textbf{\#Entity type}        & 5         & 7         & 6         & 7          & 5         & 7         & 6          & 7          \\
    \textbf{\#Relation type}      & 6         & 7         & 9         & 11         & 6         & 7         & 9          & 11         \\
    \cmidrule(l{.3em}r{.3em}){1-1} \cmidrule(l{.3em}r{.3em}){2-3} \cmidrule(l{.3em}r{.3em}){4-5} \cmidrule(l{.3em}r{.3em}){6-7} \cmidrule(l{.3em}r{.3em}){8-9}
    \textbf{\#User-entity relation}       & 198,475   & 3,431,131 & 6,862,961 & 8,548,645  & 278,280   & 4,268,838 & 8,364,580  & 11,416,146 \\
    \textbf{\#Item-entity relation}        & 2,230,660 & 2,449,630 & 2,230,660 & 2,449,630 & 3,465,140 & 3,490,318 & 3,465,140 & 3,490,318 \\
    \bottomrule
    \end{tabular}
}
\label{tab:dataset}
\end{table*}

\begin{table*}[t]
\centering
\caption{Overall performance across various knowledge graphs. 
\textnormal{The best performance is shown in \textbf{bold} for each comparison.}}
\vspace{-0.6em}
\resizebox{0.95\textwidth}{!}{%
\setlength\tabcolsep{8pt}
\begin{tabular}{ccccccccc}
    \toprule
       & \multicolumn{4}{c}{\textbf{Amazon Beauty}}  & \multicolumn{4}{c}{\textbf{Amazon Clothing}} \\
    \cmidrule(l{.3em}r{.3em}){2-5} \cmidrule(l{.3em}r{.3em}){6-9}
    \textbf{Metric~ (\%)} & \textbf{NDCG@10}  & \textbf{Recall@10} & \textbf{HR@10}     & \textbf{Precision@10} & \textbf{NDCG@10}   & \textbf{Recall@10}  & \textbf{HR@10}    & \textbf{Precision@10} \\
    \cmidrule(l{.3em}r{.3em}){1-1} \cmidrule(l{.3em}r{.3em}){2-5} \cmidrule(l{.3em}r{.3em}){6-9}
    \textbf{\(\mathcal{G}_\textrm{base}\)}  & 2.518 & 3.934  & 7.812  & 0.953     & 0.827  & 1.609   & 2.390  & 0.246     \\
    \textbf{\(\overline{\mathcal{G}}_\textrm{base}\)}  & \textbf{3.806} & \textbf{5.616}  & \textbf{10.221} & \textbf{1.180}      & \textbf{2.236}  & \textbf{3.672}   & \textbf{5.428} & \textbf{0.558}     \\
    \cmidrule(l{.3em}r{.3em}){1-1} \cmidrule(l{.3em}r{.3em}){2-5} \cmidrule(l{.3em}r{.3em}){6-9}
    \textbf{\(\mathcal{G}_\textrm{large}\)}  & 5.449 & 8.324  & 14.401 & 1.707     & 2.858  & 4.834   & 7.020  & 0.728     \\
    \textbf{\(\overline{\mathcal{G}}_\textrm{large}\)} & \textbf{5.517} & \textbf{8.560}   & \textbf{14.752} & \textbf{1.763}     & \textbf{3.026}  & \textbf{5.105}   & \textbf{7.378} & \textbf{0.765}    \\
    \bottomrule
\end{tabular}
}
\label{tab:experiment_result}
\end{table*}

\subsection{Experimental Settings}

\subsubsection*{Dataset}

We evaluate our consistent methodology on two datasets to demonstrate its applicability across both general domains and recommender system scenarios. 

\textbf{Amazon Beauty and Clothing\footnote{https://jmcauley.ucsd.edu/data/amazon/}} are e-commerce datasets from Amazon, offering diverse metadata such as brand, category, price, and other relevant attributes. Additionally, they provide information on related items connected through predefined relationships like \textsf{\textit{``also\_bought''}}, \textsf{\textit{``also\_viewed''}}, and \textsf{\textit{``bought\_together''}}, as well as user reviews for the items.

\subsubsection*{Baseline \& Knowledge graph Construction} 
PGPR \cite{xian2019pgpr} is a knowledge graph-based recommender model that leverages reinforcement learning for path-based reasoning and effectively captures relations between users and items.
We evaluated the performance of our proposed method using PGPR as the recommender model.

For our experiments, we followed the training and evaluation methods of PGPR and constructed several knowledge graphs to compare the effectiveness of incorporating context information --- \(\mathcal{G}_\textrm{base}\), \(\overline{\mathcal{G}}_\textrm{base}\), \(\mathcal{G}_\textrm{large}\), and \(\overline{\mathcal{G}}_\textrm{large}\). A brief description of each knowledge graph is provided below:
\begin{itemize}[leftmargin=1.1em]
\item \(\mathcal{G}_\textrm{base}\): A base knowledge graph constructed using item type, brand, related products, and user purchase history as side information.
\item \(\overline{\mathcal{G}}_\textrm{base}\): An enhancement of \(\mathcal{G}_\textrm{base}\) by adding general and specific topics extracted from item descriptions and user reviews.
\item \(\mathcal{G}_\textrm{large}\): An extension of \(\mathcal{G}_\textrm{base}\) by utilizing all words from the review data as entities, connected to users and items following the PGPR approach.
\item \(\overline{\mathcal{G}}_\textrm{large}\): An integrated version of \(\mathcal{G}_\textrm{large}\) with two types of topics generated from our approach.
\end{itemize} 
Detailed metadata on each knowledge graph is provided in Table \ref{tab:dataset}.

\subsubsection*{Evaluation Metrics}
The performance of the model was evaluated using four metrics: Normalized Discounted Cumulative Gain (NDCG)@10, Recall@10, Precision@10, and Hit Ratio (HR)@10. 
NDCG evaluates the quality of the predicted ranking in the recommender system by considering the relevance and position of recommended items.
Recall calculates the proportion of items that the user interacted with out of the recommended list, while Precision indicates the proportion of recommended items that the user engaged with.
Lastly, HR measures whether at least one item that the user interacted with appears in the recommended list.

\subsubsection*{Implementation Details.} 
We employed the Adam optimizer for training, with the number of epochs set to 30 and the learning rate fixed at 0.001, implemented in PyTorch. 
The batch size was set to 64, and the pruned action space is set to 200.
Since \(\mathcal{G}_\textrm{large}\) is the knowledge graph used in PGPR, we fully applied PGPR's settings without modification. We utilized a widely adopted LLM, specifically \texttt{gpt-4o-mini}, to extract general and specific topics.

\subsection{Experimental Results\label{sec:result}}

This section presents the experimental results of the proposed approach, as summarized in Table \ref{tab:experiment_result}.

We first compare \(\mathcal{G}_\textrm{base}\) and \(\overline{\mathcal{G}}_\textrm{base}\) to observe the impact of incorporating context information.
The results show that the knowledge graph integrating both general and specific topics (\(\overline{\mathcal{G}}_\textrm{base}\)) outperformed the baseline knowledge graph (\(\mathcal{G}_\textrm{base}\)), which was constructed using only side information. 
This demonstrates that combining context information with side information has a significant impact on improving recommendation performance.

We next compared the recommendation results of two extended knowledge graphs following the PGPR approach; the first (\(\mathcal{G}_\textrm{large}\)) utilized only review data, while the second (\(\overline{\mathcal{G}}_\textrm{large}\)) incorporated item descriptions in addition to reviews. 
The knowledge graph enriched with item descriptions and user reviews demonstrated superior performance, as general topics represent the broader characteristics of items, while specific topics capture detailed attributes and user preferences.
Both types of topics played a crucial role in enhancing performance, emphasizing the importance of analyzing diverse forms of side and context information.

In short, our approach can be applied across different domains, resulting in consistent performance improvements. 
These implications highlight its interoperability and effectiveness in enhancing knowledge graphs when integrated with existing methods.

\section{Conclusion}
\label{sec:06_conclusion}

This study proposes a method for systematically and consistently extracting general and specific knowledge embedded in an item's side information and context information using LLMs. 
Specifically, we extract general topics using side and context information. Specific topics are derived from the context information of each item, capturing both the objective characteristics of the items and the subjective preferences of users.
This allows us to present a knowledge graph expansion method that enhances both item-entity connections and user-entity interactions. 
Furthermore, we refine and group synonymous words using our proposed algorithm and LLM, enhancing the organic connections between entities within the knowledge graph.
Importantly, we demonstrate that this method can effectively expand the knowledge graph without requiring intervention from domain experts.

We further aim to enhance interoperability and improve recommendation performance in real-world systems by applying our method to integrated systems across multiple domains.

\section*{Acknowledgements}
This work was supported in part by the National Research Foundation of Korea (NRF) grant funded by the Korea government (MSIT) (No. NRF-2022R1C1C1012408), in part by Institute of Information \& communications Technology Planning \& Evaluation (IITP) grants funded by the Korea government (MSIT) (No. 2022-0-00448/RS-2022-II220448, Deep Total Recall: Continual Learning for Human-Like Recall of Artificial Neural Networks, and No. RS-2022-00155915, Artificial Intelligence Convergence Innovation Human Resources Development (Inha University)), and in part by the INHA UNIVERSITY Research Grant.

\balance
\bibliographystyle{ACM-Reference-Format}
\bibliography{sac25.bib} 

\end{document}